\documentclass[aps,prl,showpacs,superscriptaddress,twocolumn]{revtex4}
\usepackage{amssymb}

\usepackage[dvips]{color}
\usepackage[dvips]{epsfig}
\usepackage{amsmath}
\usepackage{bm}

\begin{document}

\title{Ferromagnetic behavior in magnetized plasmas}
\author{Gert Brodin and Mattias Marklund}
\affiliation{Department of Physics, Ume{\aa} University, SE--901 87 Ume{\aa}, Sweden}

\received{September 20, 2007}

\begin{abstract}
We consider a low-temperature plasma within a newly developed MHD Fluid
model. In addition to the standard terms, the electron spin, quantum
particle dispersion and degeneracy effects are included. It turns out that
the electron spin properties can give rise to Ferromagnetic behavior in
certain regimes. If additional conditions are fulfilled, a homogenous
magnetized plasma can even be unstable. This happen in the low-temperature
high-density regime, when the magnetic properties associated with the spin
can overcome the stabilizing effects of the thermal and Fermi pressure, to
cause a Jeans like instability.
\end{abstract}

\maketitle

Already in the 1960's, Pines studied the excitation spectrum of quantum
plasmas \cite{Pines}, for which we have a high density and a low temperature
as compared to normal plasmas. In such systems, the finite width of the
electron wave function makes quantum tunnelling effects crucial, leading to
an altered dispersion relation. Since the pioneering paper by Pines, a
number or theoretical studies of quantum statistical properties of plasmas
has been done (see e.g. Ref. \cite{kremp-etal} and references therein). For
example, Bezzerides \& DuBois presented a kinetic theory for the quantum
electrodynamical properties of nonthermal plasmas \cite{bezzerides-dubois},
while Hakim \& Heyvaerts presented a covariant Wigner function approach for
relativistic quantum plasmas \cite{hakim-heyvaerts}. Recently there has been
an increased interest in the properties of quantum plasmas, e.g. \cite
{Manfredi2005,haas-etal1,haas,shukla,garcia-etal,marklund-brodin,brodin-marklund,BM-pairplasma}%
. The studies has been motivated by the recent development in nanostructured
materials \cite{craighead} and quantum wells \cite{manfredi-hervieux}, the
discovery of ultracold plasmas \cite{li-etal} (see Ref. \cite{fletcher-etal}
for an experimental demonstration of quantum plasma oscillations in Rydberg
systems), astrophysical applications \cite{harding-lai}, or a general
theoretical interest. Moreover, it has recently been experimentally shown
that quantum dispersive effects are important in inertial confinement
plasmas \cite{glenzer-etal}. The list of quantum mechanical effects that can
be included in a fluid picture includes the dispersive particle properties
accounted for by the Bohm potential \cite
{Manfredi2005,haas-etal1,haas,shukla,garcia-etal}, the zero temperature
Fermi pressure \cite{Manfredi2005,haas-etal1,haas,shukla,garcia-etal}, spin
properties \cite{marklund-brodin,brodin-marklund,BM-pairplasma} as well as
certain quantum electrodynamical effects \cite
{Lundin2007,lundstrom-etal,Brodin-etal-2007}. Within such descriptions \cite
{Manfredi2005,haas-etal1,haas,shukla,garcia-etal,marklund-brodin,brodin-marklund,Lundin2007,lundstrom-etal,Brodin-etal-2007}
, quantum and classical collective effects can be described within a unified
picture.

In the present paper we will make use of general equations for spin plasmas
that were derived in Ref. \cite{marklund-brodin}, and further developed
towards the MHD regime in \cite{brodin-marklund}. Exploring the basic set of
equations presented in Ref.\ \cite{brodin-marklund} we demonstrate that the
standard plasma behavior can be significantly changed by the electron spin
properties, and that the plasma can even show Ferromagnetic behavior in the
low temperature high density regime. Furthermore, a homogeneous magnetized
plasma can actually be unstable, also when the spin degree of freedom is in
thermodynamic equilibrium. The instability is due to the magnetic attraction
of spins, and the mechanism is conceptually similar to the well-known Jeans
instability \cite{Jeans-instability}. Applications of our results to
laboratory and astrophysical plasmas are discussed.

Adopting the spin-MHD equations put forward in Ref. \cite{brodin-marklund},
our plasma is described by the continuity equation
\begin{equation}
\frac{\partial \rho }{\partial t}+\bm {\nabla}\cdot (\rho \bm{v})=0,
\label{eq:mhd-cont}
\end{equation}
the momentum equation
\begin{eqnarray}
&&\!\!\!\!\!\!\!\!\!\!\!\! \rho \left( \frac{\partial }{\partial t}+\bm{v}\cdot \bm{\nabla}\right) %
\bm{v}=-\bm{\nabla}\left( \frac{B^{2}}{2\mu _{0}}-\bm{M}\cdot \bm{B}\right)  -\bm {\nabla}P
\nonumber \\ &&\!\!\!\!
+ \bm{B}\cdot \bm {\nabla}\left( \frac{1}{\mu_0}\bm{B}-\bm{M}\right) - \frac{%
\hbar^2\rho}{2m_em_i}\bm{\nabla}\left( \frac{\nabla ^{2}\sqrt{\rho}}{\sqrt{%
\rho}}\right)\, ,  \label{eq:mhd-mom2}
\end{eqnarray}
and the idealized Ohm's law
\begin{equation}
\frac{\partial \bm{B}}{\partial t}=\bm{\nabla}\times \left( \bm{v}\times %
\bm{B}\right)  \label{Eq:Ohm}
\end{equation}
where $\rho $ is the plasma density, $\bm{v}$ the fluid velocity, $\bm{B}$
the magnetic field, $\bm{M}$ the magnetization, $P$ the pressure, $m_{e}$, ($m_{i}$)
denotes the electron (ion) mass and $\hbar $ is Planck%
\'{}%
s constant. In addition to the standard ideal MHD momentum equation, Eq. (%
\ref{eq:mhd-mom2}) contains the quantum Bohm potential (which tends to
smooth the density profile), as well as magnetization effects due to the
electron spin. Equations (\ref{eq:mhd-cont})--(\ref{Eq:Ohm}) should be
complemented by an expression for the Magnetization, as well as an equation
of state for the pressure. In thermodynamic equilibrium, the degree of spin
alignment with an external magnetic field is described by the Brillouin
functions $B_{s}$, where the index s is the spin number. For spin-$\frac{1}{2%
}$ particles we have $B_{1/2}(\mu _{B}B/T)=\tanh (\mu _{B}B/T)$, leading to
a corresponding Magnetization
\begin{equation}
\bm{M}=\frac{\mu _{B}\rho}{m_i}\,\tanh \left( \frac{\mu _{B}B}{T}\right)
\widehat{\bm{B}}.  \label{Eq: tanh-factor}
\end{equation}
Here $B$ denotes the magnitude of the magnetic field and $\widehat{\bm{B}}$
is a unit vector in the direction of the magnetic field, $\mu _{B}=e\hbar
/2m_{e}$ is the Bohr magneton, $e$ is the magnitude of the elementary
charge, and $T$ is the temperature given in energy units. In general the
argument of the $\tanh $-function can vary, if for example the magnetic
field strength varies. However, in case the variations of the magnetic field
occurs on a time-scale shorter than the characteristic spin relaxation time,
the degree of alignment can be considered as approximately constant. Since
spontaneous spin changes does not occur for single electrons (due to angular
momentum conservation), this spin relaxation time is not smaller than the
inverse collision frequency, which can be taken as infinite in many
applications. This case will be considered for the reminder of this article,
and thus $\tanh (\mu _{B}B/T)\rightarrow \tanh (\mu _{B}B_{0}/T_{0})$ in Eq.
(\ref{Eq: tanh-factor}), where the index $0$ denotes the unperturbed
background value. Furthermore, for low electron temperatures, it is
necessary to include the contribution from the zero temperature Fermi
pressure in the equation of state. Writing the equation of state as
\begin{equation}
\nabla P=c_{s}^{2}\nabla n
\end{equation}
we emphasize that the ion-acoustic velocity $c_{s}$ includes the
contribution from the ion and electron thermal motion, as well as the
electron Fermi pressure. Thus we have
\begin{equation}
c_{s}^{2}=v_{ti}^{2}+\frac{m_{e}}{m_{i}}\left( v_{te}^{2}+\frac{3}{5}%
v_{Fe}^{2}\right)  \label{Eq:Acoustic-velocity}
\end{equation}
where $v_{ti}$ and $v_{te}$ are the (effective) ion and electron thermal
velocities \cite{Effective-note},whereas $v_{Fe}$ is the electron 
Fermi velocity \cite{Fermi-note}. Equations (\ref
{eq:mhd-cont})--(\ref{Eq:Acoustic-velocity}) constitute a closed set that
describe the spin modified quantum MHD equations.

In what follows, we will study the linear modes of this system, with a
particular focus on the stability properties. With $n=n_{0}+n_{1}$, $\bm{B}=%
\bm{B}_{0}+\bm{B}_{1}$, $\bm{M}=\bm{M}_{0}+\bm{M}_{1}$, and $\bm{v} = \bm{v}%
_1$, such that $n_1 \ll n_0$, $|\bm{B}_1| \ll |\bm{B}_1|$, and $|\bm{M}_1|
\ll |\bm{M}_0|$, we linearize our equations in the perturbed variables.
Assuming that the background quantities are constants, the general
dispersion relation can, after a Fourier decomposition, be written
\begin{eqnarray}
&&\!\!\!\!\!\!\!\! \left( \omega ^{2}-k_{z}^{2}\widetilde{C}_{A}^{2}\right) \bigg[ \left(
\omega ^{2}-k^{2}\widetilde{C}_{A}^{2}-k_{x}^{2}V_{sA}^{2}(k)\right) \left(
\omega ^{2}-k_{z}^{2}V_{A}^{2}(k)\right)
\nonumber \\ &&
+k_{x}^{2}k_{z}^{2}cV_{sA}^{4}(k)%
\bigg] =0  \label{Eq:General-DR}
\end{eqnarray}
where $\widetilde{C}_{A}$ is the spin-modified Alfv\'{e}n velocity given by
\begin{equation}
\widetilde{C}_{A}=\frac{C_{A}}{\left[ 1+ (\hbar \omega
_{pe}^{2}/mc^{2}\omega _{ce}^{(0)})\tanh (\mu _{B}B_{0}/T_{0})\right] ^{1/2}}
\,\, ,  \label{Eq-Alf-spin-velocity}
\end{equation}
$C_{A}$ is the standard Alfv\'{e}n velocity $C_{A}=(B_{0}^{2}/\mu _{0}\rho
_{0})^{1/2}$,
\begin{equation}
V_{sA}^{2}(k)=V_{A}^{2}(k)-\frac{\hbar \omega _{ce}}{m_{i}}\tanh \left(
\frac{\mu _{B}B_{0}}{T_{0}}\right) \, ,
\end{equation}
and
\begin{equation}
V_{A}^{2}(k)=c_{s}^{2}+\frac{\hbar ^{2}k^{2}}{4m_{i}m_{e}}
\end{equation}
Here $\omega _{pe}=(n_{0}e^{2}/\varepsilon _{0}m)^{1/2}$ is the plasma
frequency, $\omega _{ce}^{(0)}$ is the electron cyclotron frequency associated with
the external magnetic field (i.e. with the contribution to $B_{0}$ from the
spin sources excluded). The relation between the full electron
cyclotron frequency $\omega _{ce}=eB_{0}/m_e$ and
$\omega _{ce}^{(0)}$ is given by $\omega_{ce}=\omega _{ce}^{(0)} + \hbar
\omega _{pe}^{2}\tanh (\mu _{B}B_{0}/T_0)/mc^{2}$. We stress that $V_{sA}$,
which to some extent can be considered as an effective acoustic velocity,
may be imaginary for a strongly magnetized plasma due to the spin
contribution, a fact which will be explored in some detail below.

In deducing Eq. (\ref{Eq:General-DR}) we have assumed that the spin
orientation has reached the thermodynamic equilibrium state in response to
the external magnetic field. This ensures there is no free energy stored
into the spin degree of freedom, and as a consequence it turns out that the
shear Alfv\'{e}n mode described by the first factor of (\ref{Eq:General-DR})
is always stable, since clearly $\widetilde{C}_{A}$ is always real. This is related to the fact
that this particular mode has no density perturbations. By
contrast the second factor, describing the fast and slow magnetosonic modes,
does not necessarily predict stability. The reason is that the electrons
carry spin, and that they thus behave as single magnets to some extent. Just
like magnets or gravitating matter, the electrons may thus attract each
other, leading to an exponentially growing density, similar to the
gravitational Jeans instability. Naturally electrostatic repulsion among the
electrons could in principle act as a strong counteracting force to this
scenario. However, within the low-frequency MHD limit, ions and electrons
move together, and thus the Coulomb force does not provide a stabilizing
mechanism. To shed some further light on the stability properties, we
consider propagation perpendicular to the external magnetic field, which is
the geometry which leads to instability most easily. For the case $\bm{k}=k%
\hat{\bm{x}}$, Eq. (\ref{Eq:General-DR}) reduces to
\begin{eqnarray}
&&\!\!\!\!\!\!\!\!\!\!\!\! \omega =k\Bigg[ \frac{C_{A}^{2}}{1+(\hbar \omega _{pe}^{2}/mc^{2}\omega
_{ce}^{(0)})\tanh (\mu _{B}B_{0}/T_{e0})}
\nonumber \\ && \quad\,\,
+c_{s}^{2}+\frac{\hbar ^{2}k^{2}}{%
4m_{i}m_{e}}-\frac{\hbar \omega _{ce}}{m_{i}}\tanh \left( \frac{\mu _{B}B_{0}%
}{T_{0}}\right) \Bigg] ^{1/2}  \label{Eq:Instability-DR} .
\end{eqnarray}
The condition for instability is thus that the last negative term of (\ref
{Eq:Instability-DR}) dominates over all the others. Under this assumption, we 
have depicted the growth rate as a function of $k$ in Fig. 1.
\begin{figure}
\includegraphics[width=.98\columnwidth]{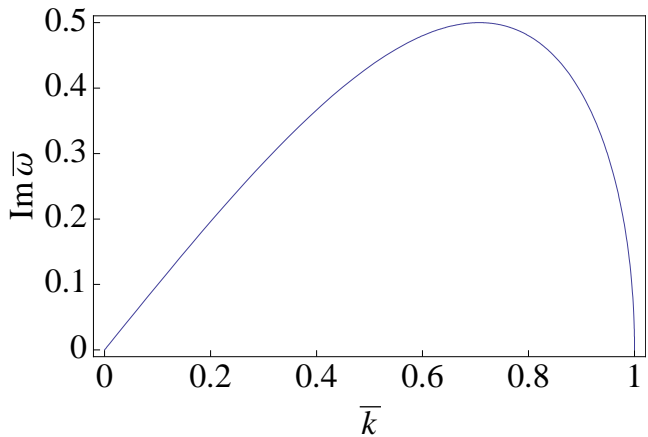}
\caption{The growth rate $\mathrm{Im}\,\bar\omega$ as a function of $\bar k$ obtained from
  the dispersion relation (11). The relation (11) is of the form $\omega = k(b^2k^2 - a^2)^{1/2}$,
  and we use the normalization $\bar\omega = |b|\omega/|a|^2$ and $\bar k = |b| k/|a|$. 
  We note that we have a maximal growth rate for the 
  wavenumber $k_{\rm max} = |a|/(\sqrt{2}|b|)=[2m_{e}[(|P_{\mathrm{sp}}|-P_{\mathrm{m}}-P)]/n_{0}\hbar ^{2}]^{1/2}$, 
  corresponding to a maximal growth rate $\gamma_{\mathrm{max}}=(m_{e}/m_{i})^{1/2}[|P_{\mathrm{sp}}|-P_{\mathrm{m}}-P)]/n_{0}\hbar$.
  Furthermore, although (11) is a linear dispersion relation, the present instability shows
  similarities to the modulational instability.
  }
\end{figure}

The necessary and sufficient instability condition can thus be written as
\begin{equation}
P_{\mathrm{sp}}+P_{\mathrm{m}}+P+\frac{n_{0}\hbar ^{2}k^{2}}{4m_{e}}<0 ,
\label{Eq:Full-condition}
\end{equation}
where the total pressure $P_{\mathrm{tot}}=P_{\mathrm{sp}}+P_{\mathrm{m}}+P$
consists of the effective spin pressure $P_{\mathrm{sp}}=-n_{0}\hbar \omega
_{ce}\tanh (\mu _{B}B_{0}/T_{0})$ which is the only negative pressure term
and therefore the source of the instability, the magnetic pressure $P_{%
\mathrm{m}}$ and the ordinary pressure $P=n_{0}m_{i}c_{s}^{2}$, containing
both the thermal and Fermi pressure part. Furthermore, the magnetic pressure
$P_{\mathrm{m}}$ is given by
\begin{equation}
P_{\mathrm{m}}=\frac{n_{0}m_{i}C_{A}^{2}}{1+\tanh (\mu _{B}B_{0}/T_{0})\hbar
\omega _{pe}^{2}/m_{e}c^{2}\omega _{ce}^{(0)}}
\label{Eq:Magn-pressure}
\end{equation}
Thus a necessary (although not sufficient) condition for instability is 
\begin{equation}
\lambda >\lambda _{c}\equiv \pi \left[ \frac{\hbar }{eB_{0}\tanh (\mu
_{B}B_{0}/T_{0})}\right]^{1/2}
\end{equation}
which means that the instability is stabilized for short wavelengths $%
\lambda =2\pi /k$, similar to the Jeans instability. The stabilizing
influence for short wavelengths stems from the Bohm potential. Furthermore,
the partial instability condition
\begin{equation}
\left| P_{\mathrm{sp}}\right| >P,  \label{Eq:Acoust-inequlity}
\end{equation}
means that a finite pressure also may lead to stabilization. \ We note from
Eq. (\ref{Eq:Acoustic-velocity}) that a low temperature is not necessary to
fulfill this condition, since the zero-temperature Fermi velocity
contributes to $c_{s}^{2}$ and thereby to $P$. However, for a sufficiently
strong magnetic field, clearly (\ref{Eq:Acoust-inequlity}) can be fulfilled.
Finally, the last subpart of the instability condition reads
\begin{equation}
\left| P_{\mathrm{sp}}\right| >P_{\mathrm{m}}  \label{Eq: Magn-ineq}
\end{equation}
which means that the magnetic pressure also acts as a stabilizer.
For a given magnetic field, this condition may be fulfilled for a
sufficiently high density. However, increasing the density means that the
Fermi velocity is increased, which may lead to a violation of (\ref
{Eq:Acoust-inequlity}). To simultaneously fulfill (\ref{Eq:Acoust-inequlity}%
) and (\ref{Eq: Magn-ineq}), and thereby to fulfill (\ref{Eq:Full-condition}),
it is required that the second term in the
denominator of the right side of (\ref{Eq:Magn-pressure}) is larger than unity,
i.e.
\begin{equation}
\omega _{ce}^{(0)}<\frac{\hbar \omega _{pe}^{2}}{m_{e}c^{2}}\tanh \left(
\frac{\mu _{B}B_{0}}{T_{0}}\right) \text{.}
\end{equation}
Since the spin cannot contribute much to the unperturbed field unless the
temperature is small enough to allow a significant alignment, this condition
in turn requires
\begin{equation}
1\lesssim \frac{\hbar ^{2}\omega _{pe}^{2}}{m_{e}c^{2}T_{0}}
\label{Eq:spin-temp}
\end{equation}
For temperatures small enough to fulfill (\ref{Eq:spin-temp}), the spin
contribution to the unperturbed field dominates over the external field, and
the plasma thus shows Ferromagnetic behavior. Contrary to a normal
ferromagnet, however, the density variations are not restricted, which
render possible the instability discussed above. However, plasmas with the
required background parameters are not easy to produce, as we can see from
the following examples: Firstly, if we chose a high density plasma like in
inertial fusion experiments, $\rho \sim 10^{9}\mathrm{kg/m}^{3}$,
ferromagnetic behavior occurs for temperatures $T\leq 10^{7}-10^{8}\mathrm{K}
$, as described by the inequality (\ref{Eq:spin-temp}). In this regime the
Alfv\'{e}n velocity can differ much from the standard Alfv\'{e}n velocity,
as given by (\ref{Eq-Alf-spin-velocity}). If, in addition, we want the
Jeans-like instability to occur, the most severe condition to fulfill is (%
\ref{Eq:Acoust-inequlity}), which require temperatures $T\leq 20\mathrm{K}$
for standard laboratory field strengths. Until a few years ago, the only
known plasmas where such low temperatures could be found were solid state
plasmas, which do not fit into the MHD-like model used here. However,
recently gaseous plasmas with ultra-low temperatures, $T\leq 10^{-3}\mathrm{K%
}$, has been constructed with the aid of Rydberg atoms \cite{li-etal,fletcher-etal}. Unfortunately, the
combined requirement of a reasonably high-density, Eq. (\ref{Eq:spin-temp}),
rules out the spin-instability described above in such a laboratory setting.

In addition to laboratory applications, the theories described above could
be adopted for astrophysical purposes \cite{harding-lai}. In magnetar atmospheres, the strong
magnetic field makes it possible to fulfill the conditions (\ref
{Eq:Acoust-inequlity}) even for a relativistic temperature. In that case we
should adopt the theory to a pair plasma \cite{BM-pairplasma} rather than an
ion-electron plasma. Furthermore, for white dwarf stars, the high density
makes the condition (\ref{Eq:spin-temp}) fulfilled.


\begin{thebibliography}{99}
\bibitem{Effective-note}  For temperatures where the thermal velocity is
much lower than the Fermi velocity, the square of the effective thermal
velocity for species $j$ ($j=i,e)$ is given by $v_{tj}^{2}=C\left( T\hbar
n^{1/3}\right) $, where $C$ is a dimensionless contant of order unity. For
temperatures much higher than the Fermi velocity we may use $%
v_{tj}^{2}=T/m_{j}$.
\bibitem{Fermi-note}  The electron Fermi-velocity is given by $v_{F}=\hbar
\left( 3\pi ^{2}n\right) ^{1/3}/m_{e}$
\end{thebibliography}
\end{document}